\documentclass[a4paper,twocolumn,11pt,accepted=2025-10-24]{quantumarticle}
\pdfoutput=1
\usepackage[utf8]{inputenc}
\usepackage[english]{babel}
\usepackage[T1]{fontenc}
\usepackage{amsmath}
\usepackage{hyperref}
\usepackage{tikz}
\usepackage{lipsum}
\usepackage{bm}
\usepackage{amsmath}
\usepackage{amssymb}
\usepackage{amsmath}
\usepackage{amsthm}
\usepackage{amsopn}
\usepackage{ulem}
\usepackage{cancel}
\usepackage{color}
\usepackage{float}
\usepackage{comment}
\usepackage{multirow}
\usepackage{bbold}
\DeclareMathOperator{\sign}{sign}

\begin{document}

\title{Non-Heisenbergian Quantum Mechanics}

\author{MohammadJavad Kazemi}
\affiliation{Department of Physics, Faculty of Science, University of Qom, Qom, Iran}
\email{kazemi.j.m@gmail.com}
\author{Ghadir Jafari}\email{gh.jafari@cfu.ac.ir}
\affiliation{Department of Physics Education, Farhangian University, P.O. Box 14665-889, Tehran, Iran}

\maketitle

\begin{abstract}
Relaxing the postulates of an axiomatic theory is a natural way to find more general theories, and historically, the discovery of non-Euclidean geometry is a famous example of this procedure. Here, we use this way to extend quantum mechanics by ignoring the \textit{heart} of Heisenberg's quantum mechanics---We do not assume the existence of a position operator that satisfies the Heisenberg commutation relation, $[\hat x,\hat p]=i\hbar$. The remaining axioms of quantum theory, besides Galilean symmetry, lead to a more general quantum theory with a free parameter $l_0$ of length dimension, such that as $l_0 \to 0$ the theory reduces to standard quantum theory. Perhaps surprisingly, this \textit{non-Heisenbergian} quantum theory, without a priori assumption of the non-commutation relation, leads to a modified Heisenberg uncertainty relation, $\Delta x \Delta p\geq \sqrt{\hbar^2/4+l_0^2(\Delta p)^2}$,  which ensures the existence of a minimal position uncertainty, $l_0$, as expected from various quantum gravity studies. By comparing the results of this framework with some observed data, which includes the first longitudinal normal modes of the bar gravitational wave detector AURIGA and the $2S-1S$ transition in the hydrogen atom, we obtain upper bounds on the $l_0$.
\end{abstract}

\section{Introduction}
There are various motivations to establish a theory of quantum mechanics without a position operator. Firstly, as shown by Pauli, there is no self-adjoint time operator conjugate to the Hamiltonian operator \cite{Pauli1958Encyclopedia,muga2007time,egusquiza1999free}, and therefore, if one considers a self-adjoint position operator conjugate to the momentum operator, space and time would not be on an equal footing---a fact that is not in harmony with special relativity\footnote{In fact,  there is no position operator that satisfies following natural demands \cite{newton1949localized,  caban2014relativistic,terno2014localization}: i) Its components be self-adjoint and commute with each other, ii) it satisfies canonical commutation relation with momentum operator, iii) It transforms like a three-vector under  rotations, and iv) It transforms consistently under Lorentz boosts. 
The operator satisfying all of the conditions but the last condition is obtained by Newton and Wigner \cite{newton1949localized}: An eigen-state of Newton-Wigner position operator in one reference frame, which represents a localized state at a point, is not localized in other reference frames. Hence, the Newton-Wigner position operator have no simple covariant meaning \cite{newton1949localized,barr2023bell}.}\cite{toller1999localization,terno2014localization,caban2014relativistic}.
Secondly, and more importantly, as shown by Hegerfeldt, strictly localized states lead to faster-than-light probability propagation \cite{hegerfeldt1974remark,hegerfeldt1980remarks,hegerfeldt1985violation}. This means that, in a self-consistent relativistic quantum mechanics, exactly localized states cannot exist \cite{thaller1984remarks,barat2003localization}. Therefore, the position operator is not a well-defined concept if the eigenstates of the position operator represent exactly localized states. Although exactly localized states are not possible, unsharp localization is possible \cite{bracken1999localizing, krekora2004relativistic,  kazemi2018probability}, which can be described using a proper positive-operator-valued measure instead of a self-adjoint position operator \cite{terno2014localization,kazemi2018probability}. In fact, even in the non-relativistic setting, recent progress regarding the Wigner-Araki-Yanase theorem shows that exactly localized states are not practically accessible via position measurements \cite{kuramochi2023wigner,busch2011position}.
Another motivation for avoiding exactly localized states comes from quantum gravity. Although a complete and agreed-upon theory of quantum gravity is still lacking, various approaches to quantum gravity \cite{hossenfelder2013minimal}, as well as \textit{gedankenexperiments} that heuristically combine quantum theory and general relativity \cite{calmet2004minimum,das2008universality,scardigli1999generalized,adler1999gravity,amelino1996limits} suggest the existence of a limit to observable spatial resolution.

The aim of the present work is to show how one can establish such a theory of quantum mechanics without a position operator.
Although we avoid the position operator, (at least unsharp) position measurements are routinely performed in quantum experiments, even in the relativistic regime \cite{sinha2010ruling, pleinert2021testing, korzh2020demonstration}. Therefore, a complete quantum theory must be able to predict the observed position distributions.
In fact, a space-time description of quantum phenomena is critical for understanding the relativistic aspects of quantum information processes \cite{caban2014relativistic, peres2004quantum} and addressing some fundamental questions in quantum theory \cite{krekora2004relativistic,wagner2011space,wagner2012local,glasgow2020space}. For example, do photons satisfy the position-momentum uncertainty relation? If so, how can it be proven without reference to a position operator \cite{terno2014localization,bialynicki2012uncertainty}?
Our starting point is the modern axiomatic framework of (generalized) quantum theory, where the observables are not restricted to self-adjoint operators. Instead, in the general case, an observable is described by a positive-operator-valued measure (POVM), which is not necessarily a projection-valued measure (PVM) corresponding to a self-adjoint operator \cite{masanes2019measurement,busch2009operational,holevo2011probabilistic,busch1989some,kunjwal2014quantum}.
A justification for this generalization is provided by Gleason's theorem \cite{busch1996quantum,busch2003quantum}, which guarantees that this structure is the most general one compatible with the probabilistic interpretation of quantum mechanics (for more details, see \cite{busch1996quantum}). Another justification, based on an operational approach to quantum theory, is given in \cite{masanes2019measurement}. In fact, the use of POVMs has been previously suggested to describe various observable distributions whose description via self-adjoint operators is problematic.  Examples include the arrival time distribution \cite{giannitrapani1997positive,hegerfeldt2010manufacturing,hegerfeldt2010symmetries}, work distribution \cite{roncaglia2014work}, screen observable \cite{werner1986screen}, among others \cite{busch2009operational}. The aim of the present work is to establish a space-resolved quantum mechanics in which the position probability distribution is given by a proper position POVM at the fundamental level. Although the use of POVMs to describe particle localization has been previously suggested \cite{busch1989some}, to the best of our knowledge, this idea has not yet been fully extended to systematically develop a complete quantum theory—including deriving its consequences for uncertainty relations, time evolution, the spectrum of interacting particle systems, and so on. Note that in conventional quantum mechanics, the position operator not only determines the position probability density but also serves other roles, especially in canonical quantization where it is used to determine the Hamiltonian operator corresponding to a classical Hamiltonian. We will discuss how a proper position POVM can also be consistently used for this purpose.

In this work, our focus is on the non-relativistic regime; however, the results shed light on the subject in the general case. The relativistic generalization will be reported in future works. In this regard, the rest of the paper is organized as follows. In Section II, we obtain the general position probability density consistent with Galilean symmetry and the general structure of quantum theory. In Section III, we show that this general position probability density gives rise to a generalized position-momentum uncertainty relation and implies a minimum position uncertainty, denoted by $l_0$. In Section IV, we discuss the time evolution in the presence of external potential and construct the associated Schr\"odinger equation without a position operator. Specifically, we study the simple harmonic oscillator in this framework and obtain an upper bound on $l_0$ by comparing the theoretical results with experimental data from the bar gravitational wave detector AURIGA. In Section V, we systematically generalize this formalism to interacting multi-particle systems, and compute the spectrum of the hydrogen atom in the presence of minimal position uncertainty. Using experimental data on the $1S-2S$ transition, we derive a more stringent upper bound on $l_0$.  In Section VI, we summarize the results and provide a discussion on future prospects.

\section{Probability density and Galilean symmetry}\label{Sec2}
According to generalized quantum mechanics \cite{busch1989some},
we assume that, for any $\bm{x}\in \mathbb{R}^3$, there is an operator $\hat{Q}_{\bm x}$ such that the position probability density at $\bm x$ is given by
\begin{equation}\label{POVM-Base}
\rho_t(\bm{x})=\langle\psi_t|\hat{Q}_{\bm x}|\psi_t\rangle .
\end{equation}
In the conventional Heisenberg quantum mechanics, $\hat{Q}_{\bm x}$ is a projection operator constructed from the eigenstates of a position operator, as $|\bm x\rangle\langle \bm x|$. We do not adopt this assumption. From the perspective of probability theory, it is not necessary for $\hat{Q}_{\bm{x}}$ to be a projection operator. Instead, it is only required that $\hat{Q}_{\bm{x}}$ be a positive operator, $\hat{Q}_{\bm{x}} \geq 0$, and that the set of $\hat{Q}_{\bm{x}}$ be normalized, i.e., $\int \hat{Q}_{\bm x} d^3x=\mathbb{1}$. These conditions make   $\{\hat{Q}_{\bm x}|\bm x\! \in \!\mathbb{R}^3\}$  a POVM.

Now, let's find the general position probability density for a free spinless particle consistent with non-relativistic space-time symmetries, including spatial and temporal translations, rotation, and Galilean boost. The space and time translation symmetries lead to
$
\hat Q^{t}_{\bm{x+a}}=e^{-i\bm{\hat{P}}.\bm{a}/\hbar}\hat Q^t_{\bm{x}}e^{i\bm{\hat{P}}.\bm{a}/\hbar}
$ and 
$
\hat Q^{t+\tau}_{\bm{x}}=e^{i\hat{H}\tau/\hbar}\hat Q^t_{\bm{x}}e^{-i\hat{H}\tau/\hbar},
$
respectively. In the infinitesimal limit, these transformations yield
\begin{eqnarray}\label{space-translation}
[\hat Q^t_{\bm{x}}, \hat{\bm{P}}]=-i\hbar\nabla \hat Q^t_{\bm{x}},
\end{eqnarray}
and
\begin{eqnarray}\label{time-translation}
[\hat Q^t_{\bm{x}}, \hat{H}] = +i\hbar\partial_t \hat Q^t_{\bm{x}},
\end{eqnarray}
where  $\hat{\bm{P}}/\hbar$ and $\hat{H}/\hbar$ are generators of space and time translations, respectively, and $\hat Q^t_{\bm{x}}$ is the time-dependent position POVM in the Heisenberg picture, defined as $\hat Q^t_{\bm{x}}\equiv e^{i\hat{H}t/\hbar}\hat{Q}_{\bm{x}}e^{-i\hat{H}t/\hbar}$.
Equation~(\ref{space-translation}) encapsulates the infinitesimal translational symmetry in our framework, analogous to the canonical commutation relation in standard quantum theory\footnote{Both Eq.~(\ref{space-translation}) in our model and the position–momentum commutation relation in standard quantum mechanics originate from (infinitesimal) spatial translation symmetry \cite{ballentine2014quantum,weinberg2015lectures}. In fact, Eq.~(\ref{space-translation}) leads to the standard commutation relation, $[\hat{X}_i,\hat{P}_j]=i\hbar\delta_{ij}$, provided that $\hat{\bm{X}}$ is defined as $\int \bm{x}\, \hat{Q}_{\bm{x}}\, d^3x$. However, $\hat{\bm{X}}$ coincides with the conventional position operator only when exactly localized states $|\bm{x}\rangle$ exist and $\hat{Q}_{\bm{x}}$ is defined as the projector $\hat{Q}_{\bm{x}}=|\bm{x}\rangle\langle \bm{x}|$. In the general case, when $\hat{Q}_{\bm{x}}$ is not a projector, $\hat{\bm{X}}$ cannot be interpreted as a position operator in the usual sense, since $\langle \bm{x}^2\rangle \equiv \int \rho(\bm{x})\, \bm{x}^2\, d^3x \neq \langle \psi | \hat{\bm{X}}^2 | \psi \rangle$.}.
Using space and time translation symmetries, $\hat{Q}^t_{\bm{x}}$ could be written in momentum representation as:
$$
\hat{Q}^t_{\bm{x}}\!=\!\!\int\!\! \! \ Q(\bm{p},\bm{k})  e^{\frac{i}{\hbar}(\bm{p}-\bm{k}).\bm{x}} \! \left[e^{i\frac{\hat{H}t}{\hbar}}|\bm{k}\rangle\langle \bm{p}| e^{-i\frac{\hat{H}t}{\hbar}}\right]\! d^3\bm{p} d^3\bm{k},
$$ 
where $Q(\bm{p},\bm{k})=\langle \bm{k}|\hat{Q}^0_{\bm{0}}|\bm{p}\rangle$.
Similarly, the Galilean boost and rotation symmetries lead, respectively, to the following commutation relations:
\begin{eqnarray}\label{boost}
[\hat Q^t_{\bm{x}}, \hat{\bm{K}}] = -i\hbar\ t \nabla \hat Q^t_{\bm{x}},
\end{eqnarray}
\begin{eqnarray}\label{rotation}
[\hat Q^t_{\bm{x}}, \hat{\bm{J}}] = -i\hbar\ \bm{x}\!\times\!\nabla\hat Q^t_{\bm{x}}.
\end{eqnarray}
Here, $\hat{\bm{K}}/\hbar$ and $\hat{\bm{J}}/\hbar$ denote the generators of the boost and rotation transformations in Hilbert space. Then, using the Lie algebra of the projective representation of the Galilean group \cite{weinberg1995quantum,ballentine2014quantum}, the equations (\ref{rotation}) and (\ref{boost}) lead to the following constraints on the function $Q(\bm{p},\bm{k})$:
\begin{eqnarray}
(\nabla_{\bm{p}}\!&+&\!\nabla_{\bm{k}})Q(\bm{p},\bm{k})=0,\nonumber\\
\nonumber \\ 
(\bm{p}\times\nabla_{\bm{p}}\!&+&\!\bm{k}\times\nabla_{\bm{k}})Q(\bm{p},\bm{k})=0,\nonumber
\end{eqnarray}
which together ensure that $Q(\bm{p},\bm{k})$ is a function of  $|\bm{p}-\bm{k}|$---The mathematical details of this derivation are provided in Appendix \ref{AppA}. Therefore, the general bilinear position probability density consistent with the non-relativistic space-time symmetries expressed in Eqs. (\ref{space-translation})–(\ref{rotation}) is given by:
\begin{equation}\label{POVM}
\rho_t(\bm{x})\!=\! \! \int \! f(|\bm{p}-\bm{k}|) \tilde{\psi}_t(\bm{p}) \tilde{\psi}_t^*(\bm{k})e^{\frac{i}{\hbar}(\bm{p}-\bm{k}).\bm{x}} d^3\bm{p} d^3\bm{k},
\end{equation}
where $\tilde{\psi}_t(\bm{p})$ is the momentum-space wave function, $\langle \bm{p}|\psi_t\rangle$, and $f(|\bm{p}-\bm{k}|)$ is an isotropic positive-definite kernel. The kernel $f(|\bm{p}-\bm{k}|)$ is also known as a radial positive definite function. The characterization of radial positive definite functions is a classical result of I. Schoenberg \cite{schoenberg1938metric}: 
A function $f(r)$ is radial positive definite on $\mathbb{R}^n$ if and only if there exists a finite positive measure $\alpha$ on $\mathbb{R}^+$ such that
$$
f(r)=\int_0^\infty \Omega_n(r u) d\alpha(u),
$$
where 
$
\Omega_n(r)=\Gamma(n/2) (2/r)^{(n-2)/2} J_{(n-2)/2}(r),
$
and $J_n$ is the Bessel function of order $n$ (see \cite{wells2012embeddings,gneiting1999radial,wendland2004scattered} for further details).

It is straightforward to verify that the probability density given in (\ref{POVM}) is conserved; it satisfies the continuity equation,
$
\partial \rho /\partial t+\nabla.\bm{J}=0,
$
where the probability  current $\bm{J}$ is given by
\begin{equation}\label{J-def}
\bm{J}\!=\!\!\int\! \frac{\bm{p}+\bm{k}}{2m} f(|\bm{p}-\bm{k}|)  \tilde{\psi}(\bm{p}) \tilde{\psi}^*(\bm{k}) e^{\frac{i}{\hbar}(\bm{p}-\bm{k}).\bm{x}} d^3\bm{p} d^3\bm{k},
\end{equation}
for a free particle with $\hat{H}=\hat{\bm{P}}^2/2m$. It should be noted that the only difference between the equations (\ref{POVM}) and (\ref{J-def}) and their counterparts in conventional quantum theory is the appearance of the kernel $f(|\bm{p}-\bm{k}|)$.

\section{Modified uncertainty relation}\label{Modified uncertainty relation}
 To facilitate a detailed comparison with conventional quantum mechanics, let us compute the moments of the position probability distribution, $\langle x_i^n \rangle=\int x_i^n \rho(\bm{x}) d^3\bm{x}$. Using the identity 
$\int  e^{i(\bm{p}-\bm{k}).\bm{x}/\hbar} d^3x=(2\pi\hbar)^3\delta(\bm{p}-\bm{k}),$
and  Eq.(\ref{POVM}) we find that,
$$
\langle x_i^n \rangle \!=\!(2\pi\hbar)^3\sum_{j=0}^n  \frac{n! (i\hbar)^j}{(n-j)!j!}\xi_j\!\!\int \! x_i^{(n-j)}|\psi(\bm{x})|^2 d^3x,
$$
where $\xi_j\equiv \frac{\partial^j}{\partial p^j_i}f(|\bm{p}-\bm{k}|)|_{\bm{p}=\bm{k}}$, and $\psi(\bm{x})$ is defined using momentum wave function as $\psi (\bm{x})\equiv (2\pi \hbar)^{-3/2}\int \tilde{\psi}(\bm{p})e^{i\bm{x}.\bm{p}/\hbar}d^3p$, as usual. Note, however, that  $\psi(\bm{x})$ should not be interpreted as a "\textit{wave function}" in the usual sense, because without a position operator and  position eigenstates, $\langle \bm{x}|\psi\rangle$ cannot be defined. Moreover, in the general case, the position probability density is given by Eq. (\ref{POVM}), not $|\psi(\bm{x})|^2$.

The normalization condition, $\int \rho(\bm{x}) d^3x=1$, ensures that $\xi_0= (2\pi \hbar)^{-3}$. Moreover, using the fact that $f$ is a positive kernel, it is easy to prove that  $f(x)\leq f(0)$ \cite{wells2012embeddings}, which ensures that $\xi_1=0$ and $\xi_2\leq 0$. Therefore, if $l_0$ is defined as $l_0=\sqrt{-\xi_2(2\pi\hbar)^3\hbar^2}$, it is straightforward to see that 
\begin{eqnarray} 
\langle x_i \rangle &=&\int x_i |\psi(\bm{x})|^2  d^3\bm{x},\label{2} \nonumber \\
\langle x^2_i \rangle &=&l_0^2+\int x^2_i |\psi(\bm{x})|^2  d^3\bm{x}, \label{3} \nonumber 
\end{eqnarray}
where $l_0$ is a positive constant with dimensions of length.
Therefore, it is clear that,
\begin{equation} \label{minimal}
(\Delta x_i)^2=l_0^2+ \int (x_i-\langle x_i \rangle)^2 |\psi(\bm{x})|^2d^3x,
\end{equation}
in which $\Delta x_i=\sqrt{\langle x_i^2\rangle-\langle x_i\rangle^2}$.  Note that, according to the fact that the quantity $\int (x_i-\langle x_i\rangle)^2|\psi(\bm{x})|^2 d^3x$  can be arbitrarily small, the equation (\ref{minimal}) ensures that
$(\Delta x_i)_{\text{Inf}}=l_0$, 
where $(\Delta x_i)_{\text{Inf}}=\text{Inf} \lbrace\Delta^{\psi} x_i\vert \ |\psi\rangle\in\mathcal{H}\rbrace$.

It should be noted that the assumption of the existence of arbitrarily sharp localized states, i.e., $l_0=0$, uniquely determines the kernel $f$ to be a constant function \cite{kijowski1974time}, 
$f(|\bm{p}-\bm{k}|)= f(0)= (2\pi \hbar)^{-3}$, 
which implies that equation (\ref{POVM}) reduces to $\rho = |\psi|^2$.
Moreover, if  $f=f(0)$,  the equation (\ref{J-def}) reduces to the usual Schr\"odinger probability current, i.e. $\bm{J}=\frac{\hbar}{m} \Im [\psi^* \nabla \psi]$. In fact, in this case, the current formalism reduces to the standard Heisenberg quantum theory.

The existence of a minimum position uncertainty modifies the usual position-momentum uncertainty relation.
Using properties of Fourier transformation, we know that
$$
\int \! (x_i-\langle x_i \rangle)^2 (p_i-\langle p_i \rangle)^2 |\psi(\bm{x})|^2|\tilde{\psi}(\bm{p})|^2 d^3xd^3p \! \geq \! \frac{\hbar^2}{4},\ \
$$
which, together with (\ref{minimal}), leads to the following modified position-momentum uncertainty relation:
\begin{equation}\label{modifieduncertainty}
\Delta x_i \Delta p_i \geq \sqrt{\hbar^2/4+l_0^2(\Delta p_i)^2}.
\end{equation}
This modified uncertainty relation is the same as the common generalized uncertainty relation at the first order, $\Delta x_i \Delta p_i \geq \hbar/2(1+2l_0^2(\Delta p_i)^2/\hbar^2)$, which appears in a wide range of approaches to quantum gravity \cite{amati1989can,maggiore1993generalized,scardigli1999generalized,das2008universality,hossenfelder2013minimal}.
Often, such modification is seen as a manifestation of a deformed canonical commutator  \cite{maggiore1993algebraic,kempf1995hilbert,fadel2022revisiting} which has been related to a deformation of space-time symmetry \cite{fadel2022revisiting,maggiore1994quantum,bosso2023fate}.  However, it is important to emphasize that we did not assume any modified symmetry or deformed commutation relation. The general form of quantum probability density \eqref{POVM} and the modified uncertainty relation \eqref{modifieduncertainty} are derived from the Galilean symmetry. A derivation of (an extended version of) the uncertainty relation (\ref{modifieduncertainty}), employing a method similar to ours but with a slightly distinct interpretation and without incorporating constraints from non-relativistic spacetime symmetries, is provided in \cite{lake2019generalised,lake2023generalised}. See also Appendix \ref{AppB}.

In the quantum gravity literature, it is often assumed that the value of $l_0$ is near the Planck scale ($l_p = 10^{-35}$ m). Nonetheless, it is worth mentioning that some theories with varying gravitational coupling, such as those with additional dimensions \cite{arkani1998hierarchy,antoniadis1998new,hossenfelder2003signatures}, imply a larger \textit{effective} Planck length. In our framework, $l_0$ is a free parameter, and we do not presuppose any specific value for it. In principle, the value of $l_0$ should be determined from a direct comparison of the theoretical result with experimental data \cite{pikovski2012probing}.
 
\section{Single particle in an external potential}
To find an experimental bound on $l_0$, we need to extend the model to non-free particles. Firstly, we consider a single particle in the presence of an external potential and then the model is extended for an interacting many-particle system.
For a particle in an external potential, we assume that the position probability density is given by a POVM as  $\rho_t(\bm{x})=\langle\psi_t|\hat{Q}_{\bm x}|\psi_t\rangle$, and the time evolution of the state is governed by the Schr\"odinger equation:  
\begin{equation}\label{Sch}
i \hbar \frac{d}{dt} |\psi\rangle = \hat{H} |\psi\rangle,
\end{equation}  
where  
$\hat H=\hat{\bm{P}}^2/2m+\hat V$,  
and the potential operator, associated with the classical scalar potential \(V(\bm{x})\), is naturally defined by  
\begin{equation}\label{Vdef}
\hat{V} = \int V(\bm{x}) \hat{Q}_{\bm{x}} \, d^3x.
\end{equation}
Note that, when $\hat{Q}_{\bm{x}}$ are projection operators, this definition reduces to the standard potential operator, i.e., $\hat{V}=\int V(\bm{x})|\bm{x} \rangle\langle \bm{x}| d^3x$.
In the general case, where $l_0 \neq 0$, this definition is conceptually consistent with our definition of the position probability density, Eq.~(\ref{POVM}), since it ensures that
\begin{equation} \label{V-expectation}
\langle \psi | \hat{V} | \psi \rangle = \int \rho^\psi(\bm{x}) V(\bm{x}) \, d^3x.
\end{equation}
In fact, the definition \eqref{Vdef}, on the other hand, can be deduced by assuming Eq.(\ref{V-expectation}).

It is easy to show that, as in the free particle case, Galilean symmetry implies
 $\langle \bm{p}| \hat{Q}_{\bm{x}}|\bm{k}\rangle=f(|\bm{p}-\bm{k}|)  e^{i(\bm{p}-\bm{k}).\bm{x}/\hbar}.$
Therefore, in the momentum representation, the modified Schr\"odinger equation (\ref{Sch}) reads 
\begin{equation}\label{Sch-momentum-potential}
i\hbar \frac{\partial}{\partial t} \tilde{\psi} (\bm{p})\!=\!\frac{\bm{p}^2}{2m}\tilde{\psi} (\bm{p})+\!\int \! \tilde{U}(\bm{p}-\bar{\bm{p}})\tilde{\psi} (\bar{\bm{p}}) d^3\bar{\bm{p}},
\end{equation}
where
$
\tilde{U}(\bm{p}-\bar{\bm{p}})=f(|\bm{p}-\bar{\bm{p}}|)\int V(\bm{x})e^{i(\bm{p}-\bar{\bm{p}}).\bm{x}/\hbar} d^3x.
$
Equations (\ref{POVM}) and (\ref{Sch-momentum-potential}) together ensure local probability conservation---they lead to the continuity equation,
$
\partial \rho /\partial t+\nabla.\bm{J}=0,
$
in which the probability current $\bm{J}$ is given by
\begin{equation}\label{J-def-potential}
\bm{J}(\bm{x})\!=\!\!\int \! \left( \alpha(\bm{p}+\bm{k})\!+\!\beta(\bm{p}-\bm{k})\right) e^{\frac{i}{\hbar}(\bm{p}-\bm{k}).\bm{x}}d^3\bm{p} d^3\bm{k},\nonumber
\end{equation}
where $\alpha = f(|\bm{p}-\bm{k}|) \tilde{\psi}^*(\bm{k})\tilde{\psi}(\bm{p})/2m$, and 
\begin{widetext}
\begin{eqnarray}
\beta =\frac{f(|\bm{p}-\bm{k}|)}{(\bm{p}-\bm{k})^2} 
\left(\!\tilde{\psi}^*(\bm{k})\!\!\int \!\! \tilde{U}(\bm{p}-\bar{\bm{p}})\tilde{\psi}(\bar{\bm{p}})d^3\bar{\bm{p}}\!-\!\tilde{\psi}(\bm{p})\!\!\int \!\! \tilde{U}^*\!(\bm{k}-\bar{\bm{k}})\tilde{\psi}^*(\bar{\bm{k}})d^3\bar{\bm{k}}\right)\!\!.\nonumber
\end{eqnarray}
\end{widetext}
It is easy to check that the given probability current transforms correctly under Galilean transformations \cite{brown1999galilean}. In principle, this probability current may be experimentally probed through measurements of the arrival time distribution
\cite{ali2003spin,ali2007quantum,vona2013does,tumulka2022detection,roncallo2023does,rafsanjani2023can,kazemi2023detection}.

The only difference between the Eq.(\ref{Sch-momentum-potential}) and the usual Schr\"odinger equation in the momentum representation is the presence of the factor $f$, which can be effectively considered as a deformation of the potential. However, it is important to remark that the physical content of this equation is different from that of the usual Schr\"odinger equation with a deformed potential, due to the presence of the factor $f$ in the definition of the position probability density, given in Equation~(\ref{POVM}). Nonetheless, the equations~(\ref{POVM}) and (\ref{Sch-momentum-potential}) together lead to
\begin{equation}\label{Ehrenfest}
m \frac{d^2 \langle \bm{x} \rangle}{dt^2} = -\langle \nabla V \rangle,
\end{equation}
where $\langle \nabla V \rangle$ is naturally defined as $\int \nabla V(\bm{x}) \rho(\bm{x}) \, d^3x$---which is exactly the same as the result in standard quantum theory.

For a more detailed comparison with standard quantum mechanics, it is useful to study the stationary states of the harmonic oscillator. For a $d$-dimensional harmonic oscillator with
$V(\bm{x})=\frac{m\omega^2}{2}\sum_i^d x_i^2$, the Hamiltonian eigenvalue problem $\hat{H}|\psi\rangle=E|\psi\rangle$, for any $f(|\bm{p}-\bm{k}|)$ kernel, leads to 
$$E\tilde\psi(\bm{p})=\!\left[\frac{\bm{p}^2}{2m}-\frac{m\omega^2}{2}\hbar^2\nabla^2_{\bm{p}}+\frac{d}{2} m l_0^2\omega^2\right]\tilde\psi(\bm{p}),$$
which is the same as the standard equation except for the last term. 
Therefore, in the momentum representation, the stationary states of the harmonic oscillator are the same as the standard ones, while the eigenenergies are given by 
\begin{eqnarray} \label{H-Energy}
E_{\bm{n}}=\sum_i^d\left(n_i\!+\!\frac{1}{2}\right)\hbar \omega +\frac{d}{2}m l_0^2 \omega^2.
\end{eqnarray}
Here, $\bm{n} = (n_1, \ldots, n_d)$ and the $n_i$ are integers. Note that although the energy eigenstates are the same as those in standard quantum theory and lead to the same momentum distribution, the position probability density differs from the usual case due to the presence of the $f$ kernel in Eq.~(\ref{POVM}). 
Moreover, from Eq.~(\ref{Ehrenfest}), it is evident that the time evolution of the position expectation value, $\langle x \rangle$, is the same as in standard quantum theory and is consistent with classical motion in the classical limit---a sharp contrast to the results found in deformed commutation relation frameworks\footnote{Using Eq.~(\ref{Ehrenfest}) for a harmonic oscillator, $m\frac{d^2\langle x \rangle}{dt^2} = -k\langle x \rangle$, it is evident that, in our model—as in standard quantum theory—the position expectation value of a harmonic oscillator follows the classical trajectory precisely. In contrast, the model based on a deformed commutation relation results in anharmonic motion \cite{bawaj2015probing, pedram2012new}, where the oscillation frequency depends on the amplitude. This is a stark contrast that, in principle, provides a possible means of experimentally distinguishing between the two models \cite{bawaj2015probing,bushev2019testing,kumar2020quantum}.
Moreover, the deformed commutation relation model leads to a different energy spectrum for a one-dimensional harmonic oscillator \cite{kempf1995hilbert,kempf1997non,chang2002exact}:
$$
E_{n}\!=\!\sqrt{1+\!\!\frac{\!m^2 l_0^4\omega^2}{4\hbar^2}}\left(n\!+\!\frac{1}{2}\right) \!\hbar \omega \!+\! \left(\! n^2\!+\! n \!+\!\frac{1}{2}\right)\frac{1}{2}m l_0^2 \omega^2,
$$
which can be used to distinguish between models.}.
Nonetheless, in our model, the higher-order moments of the position probability density,  $\langle x^m \rangle$, and their time evolution differ from those in standard quantum theory. To probe these effects, some clever experimental design is needed. For example, see the optomechanical setups proposed to investigate the effect of minimal position uncertainty near the Planck scale \cite{pikovski2012probing,kumar2018quantum,chevalier2022many}.
A detailed study of such setups in our non-Heisenbergian quantum theory is beyond the scope of the present work and is left for future studies. Nonetheless, as a first estimation, using Eq.s (\ref{H-Energy}), it is easy to show that
\begin{equation}\label{ground-state-energy}
m \omega^2 l_0^2 + \hbar \omega/2 \leq \langle p^2\rangle/2m+m\omega^2\langle x^2\rangle/2,
\end{equation}
which could be used to estimate an upper limit on $l_0$ \footnote{Note that, although the potential (and thus the Hamiltonian) is defined up to an additive constant $C$, the inequality~(\ref{ground-state-energy}) is independent of this constant. If one uses a constantly shifted Hamiltonian, $\hat{H}' = \hat{H} + C$, the Hamiltonian eigenvalues are shifted; however, the inequality~(\ref{ground-state-energy}) remains valid.}.
In Refs.~\cite{marin2013gravitational,marin2014investigation}, an experimental limit for the modal energy---i.e., the right-hand side of Eq.~(\ref{ground-state-energy})---has been reported by analyzing the residual motion of the first longitudinal mode of the AURIGA gravitational wave bar detector, which was cooled down to sub-millikelvin temperatures~\cite{vinante2008feedback}.
The measured modal energy, $ E_{\text{exp}}\!=\!1.3\!\times\!10^{-26}\text{J}$, leads to an upper bound on the minimal position uncertainty as $ l_0< 3.8 \times 10^{16} l_P$~\cite{marin2013gravitational}.

Nevertheless, it should be noted that the direct application of the one-particle theory to the center of mass of a macroscopic object with a very large mass, such as the AURIGA detector with $m=10^{13} M_p$, may be questionable---at least, the value of $l_0$ may not be the same for all objects \cite{bosso202330,chevalier2022many}. In principle, for a more accurate consideration, a multi-particle extension of our framework is needed. In the next section, we develop such an extension, and based on it, we obtain a more justifiable upper limit on $l_0$ using the observed hydrogen atom spectrum.

\section{Interacting particles system}
In this section, we extend the theory to multi-particle systems. For simplicity, we only consider a two-particle system; however, extending the results to an $N$-particle system is straightforward. As usual, the Hilbert space of the system is taken to be the tensor product of the individual one-particle Hilbert spaces. The probability density of a two-particle system in configuration space is given by
\begin{equation}\label{POVM-Base-Many}
\rho_t(\bm{x}_1,\bm{x}_2)=\langle\psi|\hat{Q}^t_{\bm{x}_1,\bm{x}_2}|\psi\rangle,
\end{equation}
where $\{\hat{Q}^t_{\bm{x}_1,\bm{x}_2}|(\bm{x}_1,\bm{x}_2)\in \mathbb{R}^6\}$ is a POVM in the two-particle Hilbert space. Similar to the single-particle theory, the non-relativistic space-time symmetries lead to the following commutation relations: 
\begin{eqnarray}
\left[\hat Q^t_{\bm{x}}, \hat{H}\right] &=& +i\hbar\partial_t \hat Q^t_{\bm{x}},\label{time-translation-Many}\\
\left[\hat Q^t_{\bm{x}}, \hat{\bm{P}}\right] &=& -i\hbar \sum_{i=1}^{N}\nabla_{\bm{x}_i} \hat Q^t_{\bm{x}},\label{space-translation-Many}\\
\left[\hat Q^t_{\bm{x}}, \hat{\bm{K}}\right] &=& -i\hbar\ t \sum_{i=1}^{N}\nabla_{\bm{x}_i} \hat Q^t_{\bm{x}},\label{boost-Many}\\
\left[\hat Q^t_{\bm{x}}, \hat{\bm{J}}\right] &=& -i\hbar\sum_{i=1}^{N} \bm{x}_i\!\times\!\nabla_{\bm{x}_i}\hat Q^t_{\bm{x}},\label{rotation-Many}
\end{eqnarray}
where $\bm{x}=(\bm{x}_1,\bm{x}_2)$. It is easy to check that the above commutation relations lead to
$
\langle \bm{k}_1,\bm{k}_2|\hat{Q_{\bm{0}}^0}|\bm{p}_1,\bm{p}_2\rangle=f(|\bm{p}_1-\bm{k}_1|,|\bm{p}_2-\bm{k}_2|)
$, 
where $f$ is a positive semi-definite kernel in $\mathbb{R}^6$. If, as in usual quantum theory, we assume that the separable multi-particle states lead to a separable two-particle probability density---i.e., $|\psi\rangle=\prod_i|\psi_i\rangle \Rightarrow \rho(\bm{x})=\prod_i\rho_i(\bm{x}_i)$---then we have
\begin{equation}\label{f}
f(|\bm{p}_1-\bm{k}_1|,|\bm{p}_2-\bm{k}_2|)=f_1(|\bm{p}_1-\bm{k}_1|)f_2(|\bm{p}_2-\bm{k}_2|),\nonumber
\end{equation}
where $f_1$ and $f_2$ are radial positive kernels in $\mathbb{R}^3$. Therefore, the two-particle probability density in configuration space is given by
\begin{widetext}
\begin{equation}\label{many-probablity}
\rho_t(\bm{x})= \int  \prod_i f_i(|\bm{p}_i-\bm{k}_i|) \ \tilde{\psi}_t(\bm{p}) \tilde{\psi}_t^*(\bm{k})\ e^{i(\bm{p}_i-\bm{k}_i).\bm{x}_i/\hbar} \ d^3\bm{p}_i  d^3\bm{k}_i, \nonumber
\end{equation}
\end{widetext}
where $\bm{x}=(\bm{x}_1,\bm{x}_2)$, $\bm{p}=(\bm{p}_1,\bm{p}_2)$, and $\bm{k}=(\bm{k}_1,\bm{k}_2)$. Note that we consider a system consisting of two different particles, which, in principle, could have different quantum behaviors characterized by different kernels.

We assume that the time evolution of the state is given by the Schr\"odinger equation,
$i\hbar \frac{d}{dt}|\psi\rangle=\hat H |\psi\rangle$, in which $\hat H=\sum_{i=1}^{2}\frac{\hat{\bm{P}}_i^2}{2m_i}+\hat V$ and the potential operator $\hat V$, associated with the classical potential $V(\bm{x}_1,\bm{x}_2)$,  is defined as follows:
$$
\hat{V}=\int V(\bm{x}_1,\bm{x}_2)\hat{Q}_{\bm{x}_1,\bm{x}_2} d^3\bm{x}_1 d^3\bm{x}_2.
$$
It should be noted that, when $\hat{Q}_{\bm{x}}$ are projection operators, this definition reduces to the standard potential operator---i.e., $\hat{V}=\int V({\bm{x}_1,\bm{x}_2})|{\bm{x}_1,\bm{x}_2} \rangle\langle {\bm{x}_1,\bm{x}_2}| d^3x_1 d^3x_2$.
The Galilean symmetry limits the possible forms of the interaction potential---from  $[\hat{\bm{K}},\hat{H}]=i\hbar \hat{\bm{P}}$ and Eq.(\ref{boost-Many}), or from $[\hat{\bm{P}},\hat{H}]=0$ and  Eq.(\ref{space-translation-Many}), it is easy to see that,   
\begin{equation}\label{V-cnstraint-1}
(\nabla_{\bm{x}_1}+\nabla_{\bm{x}_2})V(\bm{x}_1,\bm{x}_2)=0.
\end{equation}
Moreover, $[\hat{\bm{J}},\hat{H}]=0$ together with Eq.(\ref{rotation-Many}) lead to
\begin{equation}\label{V-cnstraint-2}
(\bm{x}_1\times\nabla_{\bm{x}_1}+\bm{x}_2\times\nabla_{\bm{x}_2})V(\bm{x}_1,\bm{x}_2)=0.
\end{equation}
The equations (\ref{V-cnstraint-1}) and (\ref{V-cnstraint-2}) imply that the potential must be a function of  $|\bm{x}_2-\bm{x}_1|$, which is the same condition that arises from Galilean symmetry in conventional quantum mechanics \cite{weinberg2015lectures}.

As a crucial example, we investigate the energy levels of the hydrogen atom within the introduced framework. The hydrogen atom may be considered as a two-particle system with interaction potential  $V(|\bm{x}_1-\bm{x}_2|)=-e^2/4\pi\epsilon_0|\bm{x}_1-\bm{x}_2|$.  In momentum representation, the energy eigenvalue problem $E|\Psi\rangle=\hat H|\Psi\rangle$ can be written as:
\begin{eqnarray}
&&E\tilde{\Psi}(\bm{p}_1,\bm{p}_2) =\left(\frac{\bm{p}^2_1}{2m_1} +\frac{\bm{p}^2_2}{2m_2}\right)\tilde{\Psi}(\bm{p}_1,\bm{p}_2)\\ \nonumber 
&&+ \int \tilde{U}(\bm{p}_1-\bm{k}_1,\bm{p}_2-\bm{k}_2) \tilde{\Psi}(\bm{k}_1,\bm{k}_2)d^3\bm{k}_1d^3\bm{k}_2,
\end{eqnarray}
where $\tilde{\Psi}(\bm{p}_1,\bm{p}_2)=\langle \bm{p}_1,\bm{p}_2|\Psi\rangle$, and
\begin{eqnarray}
&&\tilde{U} (\bm{p}_1,\bm{p}_2) = f_1(|\bm{p}_1|)f_2(|\bm{p}_2|)\nonumber \\
&& \times  \!\int V(|\bm{x}_1-\bm{x}_2|)e^{\frac{i}{\hbar}(\bm{p}_1.\bm{x}_1+\bm{p}_2.\bm{x}_2)}d^3\bm{x}_1d^3\bm{x}_2.\nonumber
\end{eqnarray}
It is well-known that, in the case of two interacting particles, working with the total momentum $\bm{p}_c=\bm{p}_1+\bm{p}_2$ and the relative momentum $\bm{p}_r=(m_1\bm{p}_2-m_2\bm{p}_1)/(m_1+m_2)$ is useful.
After separating the external and internal degrees of freedom, $\tilde{\Psi}(\bm{p}_1,\bm{p}_2)=\tilde{\psi}_c(\bm{p}_c)\tilde{\psi}_r(\bm{p}_r)$, the two-body problem can be split into  
$E_c\psi_c(\bm{p}_c) =\frac{\bm{p}^2_c}{2M} \psi_c(\bm{p}_c)$, and
\begin{equation}\label{relative-Sch-momentum}
E_r\tilde{\psi}_r(\bm{p}_r) =\frac{\bm{p}^2_r}{2\mu} \tilde{\psi}_r(\bm{p}_r)+\int \tilde{U}_r(\bm{p}_r-\bm{k}_r)\tilde{\psi}_r(\bm{k}_r)d^3\bm{k}_r,
\end{equation}
where $\bm{r}$ represents the relative coordinate, $\bm{x}_2-\bm{x}_1$, $\mu=m_1m_2/(m_1+m_2)$, $E=E_c+E_r$, $M=m_1+m_2$,  and 
$$\tilde{U}_r(\bm{p})=(2\pi\hbar)^3f_1(|\bm{p}|)f_2(|\bm{p}|) \int V(|\bm{x}|)e^{i\bm{p}.\bm{r}/\hbar} d^3\bm{r}.$$
Using the Taylor expansion of $f_i$, in the first-order approximation,  Eq.(\ref{relative-Sch-momentum}) reduces to
\begin{eqnarray}
 \left[E +\frac{\hbar^2}{2\mu}\!\nabla^2 +(1\!+\!\frac{l_1^2+l_2^2}{2} \nabla^2 )V(r)\right]\psi_r(\bm{r})=0, \nonumber
\end{eqnarray}
where 
$\psi_r(\bm{r})=(2\pi\hbar)^{-3/2}\int \tilde{\psi}_r(\bm{p})e^{i\bm{p}.\bm{r}/\hbar}d^3p$,
and  $l_1$ and $l_2$ are minimal position uncertainty of the particles, which are given by $l_i^2=(2\pi\hbar)^3\hbar^2f''_i(0)$.
Applying the usual perturbation theory, the leading corrections to the $S$-energy levels are given by
$$
\Delta E_{n}=-\frac{e^2}{2a_0 n^2} \frac{16\pi(l_1^2+l_2^2)}{n a_0^2},
$$
in which $a_0=4\pi\epsilon_0\hbar^2/\mu e^2$ is the Bohr radius. Comparing this result with the high-precision experimental data of hydrogen atom spectroscopy can lead to an upper bound for the minimal position uncertainties. To achieve this, we focus on the extremely narrow $1S-2S$ two-photon transition. In recent years, the frequency of the $1S-2S$ transition has been measured with high accuracy using a cesium clock with a relative uncertainty of $4.5 \times 10^{-15}$ \cite{matveev2013precision,parthey2011improved}.
This relative uncertainty leads to the following upper bound on the minimal position uncertainties:
$$
l_i\leq\sqrt{l_1^2+l_2^2}\leq 2.8\times 10^{16} l_p,
$$
which is in agreement with the upper bound obtained in the previous section using AURIGA data.

Finally, it is important to remark that the corrected energy levels in our framework differ from the ones in the deformed commutation relation models \cite{stetsko2006perturbation,bouaziz2010hydrogen,brau1999minimal, quesne2010composite}.
In the first order of perturbation, the conventional GUP model, based on deformed commutation relations, yields the following energy spectrum correction for the hydrogen atom \cite{quesne2010composite}:
\begin{eqnarray*}
\Delta E_{n} = \frac{e^2}{2a_0 n^2} &&\left[ \frac{2}{3} \frac{l_1^2 m_1 + l_2^2 m_2}{a_0^2 (m_1 + m_2)} \frac{8n - 3}{n^2} \right.\\
&&+  \left.\frac{20}{9 \hbar^2} \frac{l_1^2 m_1 + l_2^2 m_2}{m_1 + m_2} m_1 m_2 v^2 \right],
\end{eqnarray*}
where $v$ denotes the center-of-mass velocity of the hydrogen atom. In particular, unlike the result of our model, the energy spectrum of the hydrogen atom depends on its center-of-mass velocity. In principle, this difference, which arises as a consequence of Galilean symmetry breaking in the conventional GUP model, can be used to distinguish experimentally between this model and our model.

\section{Summary and out-looks}
A clear understanding of the foundational structure of quantum mechanics requires disentangling the implicit logical connections between its basic assumptions and derived results. One of the most important assumptions of Heisenberg’s quantum mechanics is the existence of a self-adjoint position operator that satisfies the canonical commutation relation. This assumption is usually considered necessary to derive most of the important results of the theory, especially the uncertainty relation. Nonetheless, in this work,  by ignoring the position operator, we demonstrate that Heisenberg quantum mechanics can be interpreted as a special case of a more general self-consistent quantum theory---one that retains most of the fundamental characteristics of Heisenberg’s formulation, including Galilean symmetry, probability conservation, the position-momentum uncertainty relation, a discrete and bounded-from-below energy spectrum for bound states, and so on.

In this work, the POVM formalism is used to describe the intrinsic minimal spatial resolution at the fundamental level, not to describe the effective resolution of the measurement apparatus, which has usually been studied in the literature. This is not merely a conceptual reinterpretation of existing formalism. Our interpretation leads to new dynamics and so distinct phenomenological predictions. Note that when POVMs are employed merely to model limitations of apparatus resolution, the underlying Hamiltonian remains unaffected. In contrast, assuming an \textit{intrinsic} minimal spatial resolution implies that the position operator is no longer well-defined at the fundamental level. As a result, the Hamiltonian must be constructed directly from the position POVM, rather than from the position operator. This leads to a modification of the dynamics—a feature absent in previous works and one that emerges naturally from our distinct physical interpretation.

It is important to stress that our modified dynamic is fundamentally different from those of conventional GUP frameworks that rely on deformed commutation relations \cite{maggiore1993algebraic,kempf1995hilbert,fadel2022revisiting}. This difference leads to distinct physical consequences at the phenomenological level. From a foundational perspective, a key distinction lies in the fact that our model is developed in full accordance with the standard non-relativistic space-time symmetries. In contrast, the framework of deformed commutation relations tends to break such symmetries \cite{fadel2022revisiting,maggiore1994quantum,bosso2023fate}—The usual canonical commutation relation is a direct result of the Galilean and also Poincare algebras if we consider a self-adjoint position operator that correctly transforms under spatial translation. Therefore, it appears that our formalism is more suitable than the deformed commutation relation approach for constructing a non-relativistic quantum theory with a non-zero minimal position uncertainty.

It is valuable to explore a natural extension of the present study. A discerning reader may question whether both the canonical position and momentum operators could be substituted with POVMs. Would such a substitution yield a quantum theory aligned with the \textit{extended} generalized uncertainty principle (EGUP)~\cite{bolen2005, bambi2008,park2008the,lake2023generalised}? In Appendix~\ref{AppB}, we address these questions in detail and demonstrate that, at least within the framework of Galilean symmetry, this approach does not yield a theory more general than the model proposed in this paper.

One of the most important extensions of the present work is its relativistic extension. This could provide a way to circumvent the fundamental problems in relativistic quantum mechanics, which arise from the impossibility of defining a proper relativistic position operator. In fact, due to the problems mentioned in the Introduction and some other issues \cite{sebens2019electromagnetism,kazemi2018probability,terno2014localization}, despite much effort in recent years \cite{kowalski2011salpeter,wagner2011space,wagner2012local,bialynicki2012uncertainty,babaei2017quantum,kazemi2018probability,kiessling2018quantum,sebens2019electromagnetism,hawton2019maxwell,hawton2021photon,glasgow2020space}, a complete and agreed-upon formulation of relativistic quantum mechanics with a clear space-time resolution has not yet been proposed.  Nonetheless, in relativistic quantum experiments,
the position measurements are routinely performed with high accuracy (for example see \cite{sinha2010ruling, pleinert2021testing, korzh2020demonstration}), and thus, a complete quantum theory must be able to predict the observed position distributions \cite{sebens2019electromagnetism,krekora2004relativistic}. We believe that a proper relativistic extension of our formulation of quantum theory without a position operator could circumvent the aforementioned problems, as has been previously shown by one of the authors in a very special case \cite{kazemi2018probability}. The detailed study of the relativistic extension is left for future works.

It should be noted that the generalization of quantum theory without a position operator is natural and interesting from the perspective of some non-standard interpretations of quantum theory, such as Bohmian mechanics \cite{Barnea2018Matter}, Nelson stochastic mechanics \cite{nelson1966derivation}, and the many interacting worlds model \cite{hall2014quantum}. In these theories, the operators are not fundamental; rather, they are secondary mathematical objects useful for computing statistically observable results \cite{durr2004quantum}.
However, in the formalism of these theories, the position probability density and probability current have fundamental roles. Therefore, the approach of the present work could shed  new light on the generalization of these alternative formulations of quantum theory, especially in the relativistic regime, which is an interesting open problem \cite{foo2022relativistic,durr2014can,gisin2018bohmian,holland1995quantum,durr2004bohmian}.  
\\

\section{Acknowledgment} 
We are grateful to  B. Taghavi and B. Askari for useful comments.

\appendix
\section{Derivation of the Functional Form of the Kernel $f$}\label{AppA}
In this appendix, we derive the constraint on \(\hat{Q}^t_{\bm{x}}\) that follows from non-relativistic space–time symmetries. As discussed in Section \ref{Sec2}, the infinitesimal transformations associated with temporal translations, spatial translations, Galilean boosts, and rotations lead, respectively, to the following commutation relations:
\begin{eqnarray}
\left[\hat Q^t_{\bm{x}}, \hat{H}\right] \!&=&\! +i\hbar \ \partial_t \hat Q^t_{\bm{x}},\label{T-trans}\\
\left[\hat Q^t_{\bm{x}}, \hat{\bm{P}}\right] \!&=&\! -i\hbar \ \nabla \hat Q^t_{\bm{x}},\label{X-trans}\\
\left[\hat Q^t_{\bm{x}},\! \hat{\bm{K}}\right] \!&=&\! -i\hbar\ t \nabla \hat Q^t_{\bm{x}},\label{boost-trans}\\
\left[\ \! \hat Q^t_{\bm{x}}, \hat{\bm{J}}\right] \!&=&\! -i\hbar\ \bm{x}\!\times\!\nabla\hat Q^t_{\bm{x}}.\label{rot-trans}
\end{eqnarray}
where \(\hat{H}/\hbar\), \(\hat{\bm{P}}/\hbar\), \(\hat{\bm{K}}/\hbar\), and \(\hat{\bm{J}}/\hbar\) denote the generators of time translations, spatial translations, Galilean boosts, and rotations, respectively, in the Hilbert space.
These operators satisfy the (centrally extended) Galilean algebra \cite{bargmann1954unitary,weinberg1995quantum, ballentine2014quantum}:
\begin{eqnarray}
\left[\hat{P}_{\alpha}, \hat{P}_{\beta}\right]&=&0,\\
\left[\hat{K}_{\alpha},\! \hat{K}_{\beta}\!\right]&=&0, \\
\left[\hat{P}_{\alpha},\ \hat{H}\right]&=& 0,\\
\left[\hat{J}_{\alpha},\ \hat{H}\right]&=& 0,\\
\left[\hat{K}_{\alpha}, \hat{H}\right]&=& i \hbar \hat{P}_{\alpha},\\
\left[\hat{K}_{\alpha}, \! \hat{P}_{\beta}\right]&=&i \hbar M \delta_{\alpha\beta},\\
\left[\ \!\!\hat{J}_{\alpha}, \ \!\hat{J}_{\beta}\right]&=& i\hbar\epsilon_{\alpha\beta\gamma}\hat{J}_\gamma,\\
\left[\!\hat{J}_{\alpha}, \ \hat{P}_{\beta}\right]&=& i\hbar\epsilon_{\alpha\beta\gamma}\hat{P}_\gamma,\\
\left[\hat{J}_{\alpha}, \hat{K}_{\beta}\right]&=& i\hbar\epsilon_{\alpha\beta\gamma}\hat{K}_\gamma,
\end{eqnarray}
where \(\alpha,\beta, \gamma\in \{1,2,3\}\). 
In the unitary irreducible representations of the centrally extended Galilean algebra, Schur’s lemma implies that the Casimir operators  
$$
M,\quad U \equiv \hat{H} - \frac{\hat{\bm{P}}^2}{2M},\quad \hat{\bm{S}} \equiv \hat{\bm{J}} - \frac{\hat{\bm{K}}\times\hat{\bm{P}}}{M}
$$
act as multiples of the identity. For a free, spinless particle of fixed mass \(m\) and vanishing spin, this uniquely fixes the forms of \(\hat{H}\) and \(\hat{\bm{J}}\) as follows \cite{bargmann1954unitary,ballentine2014quantum}:
\begin{eqnarray}
\hat{H}&=&\hat{\bm{P}}^2/2m + \varepsilon,\label{Hamiltoniandef}\\
\hat{\bm{J}}&=& \hat{\bm{K}} \times \hat{\bm{P}}/m,\label{Angularmomentumdef}
\end{eqnarray}
where the constant \(\varepsilon\) represents the internal energy, which is usually set to zero for a free particle.
Therefore, by applying spatial and temporal translation symmetries, Eqs. (\ref{T-trans})–(\ref{X-trans}), we obtain
\begin{eqnarray}\label{ApenQ}
\hat{Q}^t_{\bm{x}}=\int d^3\bm{p}d^3\bm{k} Q(\bm{p},\bm{k})e^{\frac{i}{\hbar}[(\bm{p}-\bm{k})\cdot\bm{x}-(E_{\bm{p}}-E_{\bm{k}})t]}|\bm{k}\rangle\langle \bm{p}|, \nonumber \\
\end{eqnarray}
where \(Q(\bm{p},\bm{k})=\langle \bm{k}|\hat{Q}_{\bm{0}}|\bm{p}\rangle\), and \(E_{\bm{p}}=\bm{p}^2/2m\).
By inserting Eq.~(\ref{ApenQ}) into Eqs.~(\ref{boost-trans})–(\ref{rot-trans}) and taking expectation values in an arbitrary state \(|\psi\rangle\), we obtain the following relations:
\begin{widetext}
\begin{eqnarray}\label{AppenBoost}
\int   Q(\bm{p},\bm{k}) \xi_{\bm{x}}^t(\bm{p},\bm{k}) \langle \psi|[|\bm{k}\rangle\langle \bm{p}|,\hat{\bm{K}}]|\psi\rangle d^3\bm{p}d^3\bm{k} = 
t \! \int  (\bm{p}-\bm{k}) Q(\bm{p},\bm{k}) \xi_{\bm{x}}^t(\bm{p},\bm{k})\langle \psi|\bm{k}\rangle\langle \bm{p}|\psi\rangle d^3\bm{p}d^3\bm{k},
\end{eqnarray}
and
\begin{eqnarray}\label{AppenRot}
\int   Q(\bm{p},\bm{k}) \xi_{\bm{x}}^t(\bm{p},\bm{k}) \langle \psi|[|\bm{k}\rangle\langle \bm{p}|,\hat{\bm{J}}]|\psi\rangle d^3\bm{p}d^3\bm{k} = 
\bm{x} \! \times  \! \int  (\bm{p}-\bm{k})  Q(\bm{p},\bm{k}) \xi_{\bm{x}}^t(\bm{p},\bm{k})\langle \psi|\bm{k}\rangle\langle \bm{p}|\psi\rangle d^3\bm{p}d^3\bm{k}, 
\end{eqnarray}
\end{widetext}
where \(\xi_{\bm{x}}^t(\bm{p},\bm{k})\!=\!e^{\frac{i}{\hbar}[(\bm{p}-\bm{k})\cdot\bm{x}-(E_{\bm{p}}-E_{\bm{k}})t]}\).
According to the commutation relation \([\hat{K}_i,\hat{P}_j]=im\hbar^2\delta_{ij}\), and based on the Stone–von Neumann theorem \cite{summers2001stone,stone1930linear}, the boost generator can be represented in the momentum representation as \(\hat{\bm{K}}=im \hbar \nabla_{\bm{p}}\). 
Therefore, it follows straightforwardly that
\begin{widetext}
\begin{eqnarray}
\langle \psi|[|\bm{k}\rangle\langle \bm{p}|,\hat{\bm{K}}]|\psi\rangle &=& im\hbar \left(\!\tilde{\psi}^*(\bm{k})\nabla_{\bm{p}}\tilde{\psi}(\bm{p})+\tilde{\psi}(\bm{p})\nabla_{\bm{k}}\tilde{\psi}^*(\bm{k})\!\right),\\
\langle \psi|[|\bm{k}\rangle\langle \bm{p}|,\hat{\bm{J}}]|\psi\rangle  &=& im \hbar \left( \bm{p}\!\times\!\nabla_{\bm{p}}\tilde{\psi}(\bm{p})\tilde{\psi}^*(\bm{k})+  \bm{k}\!\times\!\nabla_{\bm{k}}\tilde{\psi}^*(\bm{k})\tilde{\psi}(\bm{p})\right).
\end{eqnarray}
\end{widetext}
Hence, by applying integration by parts and assuming that the boundary terms vanish\footnote{
To ensure the vanishing of the boundary terms, it is necessary to impose certain regularity and decay assumptions on $\tilde{\psi}(\bm{p})$ and $Q(\bm{p},\bm{k})$. Specifically, if $\tilde{\psi}(\bm{p})$ belongs to the Schwartz space $\mathcal{S}(\mathbb{R}^3)$ and $Q(\bm{p},\bm{k})$ is a Schwartz kernel, then all boundary contributions from integration by parts vanish \cite{kanwal2004tempered}. We recall that $\mathcal{S}(\mathbb{R}^3)$ is dense in the Hilbert space $L^2(\mathbb{R}^3)$ and invariant under the action of the Galilei group, making it a natural common domain for the relevant observables. Therefore, we consider these additional assumptions throughout this work. While weaker conditions on $\tilde{\psi}(\bm{p})$ and $Q(\bm{p},\bm{k})$ may suffice for the boundary terms to vanish, identifying such minimal requirements is beyond the scope of the present work.},
one obtains the simplified forms of Eqs.~(\ref{AppenBoost}) and (\ref{AppenRot}) as follows:
\begin{widetext}
\begin{eqnarray}\label{AppenBoost2}
\int  \tilde{\psi}^*(\bm{k}) \tilde{\psi}(\bm{p})  \left(\nabla_{\bm{k}} Q(\bm{p},\bm{k})+\nabla_{\bm{p}} Q(\bm{p},\bm{k})\right) d^3\bm{k}d^3\bm{p}=\bm{0}, 
\end{eqnarray}
\begin{eqnarray}\label{AppenRot2}
\int  \tilde{\psi}^*(\bm{k}) \tilde{\psi}(\bm{p}) \!\left(\bm{k}\!\times\!\nabla_{\bm{k}} Q(\bm{p},\bm{k})+\bm{p}\!\times\!\nabla_{\bm{p}} Q(\bm{p},\bm{k})\right) \! d^3\bm{k}d^3\bm{p}=\bm{0}.
\end{eqnarray}
\end{widetext}
Since linear combinations of functions $\tilde{\psi}_1^*\otimes\tilde{\psi}_2$ are dense in $\mathcal{D}(\mathbb{R}^3\times \mathbb{R}^3)$, Eqs.~(\ref{AppenBoost2}) and (\ref{AppenRot2}) imply, respectively, that
\begin{eqnarray}\label{ACons1}
\nabla_{\bm{k}} Q(\bm{p},\bm{k})\!+\nabla_{\bm{p}} Q(\bm{p},\bm{k})=0, 
\end{eqnarray}
and
\begin{eqnarray}\label{ACons2}
\bm{k}\!\times\!\nabla_{\bm{k}} Q(\bm{p},\bm{k})\!+\!\bm{p}\!\times\!\nabla_{\bm{p}} Q(\bm{p},\bm{k})=\bm{0}.
\end{eqnarray}
Solving Eq.~(\ref{ACons1}) yields 
\begin{eqnarray}\label{ACons3}
Q(\bm{p},\bm{k})=\eta(\bm{p}-\bm{k}).
\end{eqnarray}
Inserting Eq.~(\ref{ACons3}) into Eq.~(\ref{ACons2}) implies
\begin{eqnarray}\label{ACons4}
\bm{z}\!\times\!\nabla_{\bm{z}} \eta (\bm{z})=\bm{0},
\end{eqnarray}
where $\bm{z}=\bm{p}-\bm{k}$.
Finally, Eq.~(\ref{ACons4}) implies that \(\eta(\bm{z})\) is a radial function,
$\eta (\bm{z})=f(|\bm{z}|)$, which means  
\begin{eqnarray}
Q(\bm{p},\bm{k})=f(|\bm{p}-\bm{k}|),
\end{eqnarray}
where $f$ is a positive radial kernel.

\section{Momentum POVM}\label{AppB}
In the model presented in the main text of this paper, we assume, as in standard quantum theory, that the generator of spatial translations, $\hat{\bm{P}}$, is the operator describing the physical momentum, corresponding to $m \bm{v}$.  
However, this interpretation does not appear to be strictly necessary. One may instead assume that, in addition to the position distribution, the physical momentum distribution is also described by an appropriate POVM as  
\begin{equation}\label{Pi-0}
\tilde{\rho}(\bm{\pi}) = \langle \psi | \hat{\Pi}_{\bm{\pi}} | \psi \rangle,
\end{equation}
where $\bm{\pi}$ denotes the physical momentum, $m \bm{v}$. 
In this Appendix, we carefully investigate this possibility and demonstrate that, although kinematically feasible, this generalization faces constraints from a dynamical perspective. Specifically, Galilean symmetry imposes limitations on the dynamics that prevent this approach from producing a model more general than the one presented in the manuscript. The details of this analysis are provided below.
\subsection{Constraints on the momentum kernel from space-time symmetries}
We begin by examining the constraints on $\hat{\Pi}_{\bm{\pi}}$ that arise from non-relativistic space-time symmetries.  Temporal translation, spatial translation, Galilean boost and rotation, imply the following commutation relations, respectively:
\begin{eqnarray}
\left[\hat{\Pi}_{\bm{\pi}},\hat{H}\right]\!\!&=&\!\! 0, \label{T-Trans-inMomentum}\\
\left[\hat{\Pi}_{\bm{\pi}},\hat{\bm{P}}\right]\!\!&=&\!\!\bm{0},\label{X-Trans-inMomentum}\\
\left[\!\hat{\Pi}_{\bm{\pi}},\hat{\bm{K}}\right]\!\!&=&\!\! i \hbar m \nabla_{\bm{\pi}} \hat{\Pi}_{\bm{\pi}},\label{Boost-inMomentum}\\
\left[\hat{\Pi}_{\bm{\pi}},\hat{\bm{J}}\right]\!\!&=&\!\! i \hbar \bm{\pi} \! \times\! \nabla_{\bm{\pi}} \hat{\Pi}_{\bm{\pi}}.\label{Rot-inMomentum}
\end{eqnarray}
In the following, we show that the above symmetry conditions determine $\hat{\Pi}_{\bm{\pi}}$ up to a \textit{single-variable} kernel.
The Galilean boost symmetry condition, whose infinitesimal form is given by Eq.~(\ref{Boost-inMomentum}), implies that  
\begin{equation}
\hat{\Pi}_{\bm{\pi}} = e^{-i \hat{\bm{K}} \cdot \bm{\pi} / \hbar m} \, \hat{\Pi}_{\bm{0}} \, e^{i \hat{\bm{K}} \cdot \bm{\pi} / \hbar m}.
\end{equation}
Therefore, using the completeness relation $\int d^3\bm{p} \, |\bm{p}\rangle\langle \bm{p}| = \mathbb{I}$, the momentum POVM can be written as:
\begin{equation}\label{Pi-1}
\hat{\Pi}_{\bm{\pi}} \!=\! \int \! d^3\bm{p} d^3\bm{k}  e^{-i \hat{\bm{K}} \cdot \bm{\pi} / \hbar m} |\bm{k}\rangle \langle \bm{k} | \hat{\Pi}_{\bm{0}} | \bm{p} \rangle \langle \bm{p} | e^{i \hat{\bm{K}} \cdot \bm{\pi} / \hbar m},
\end{equation}
where $|\bm{k}\rangle$ and $|\bm{p}\rangle$ denote the eigenstates of the translation generator, $\hat{\bm{P}}$.
Using the (centrally extended) Galilean  algebra, $\left[\hat{K}_i,\hat{P}_j\right]=im \hbar\delta_{ij} $,  Eq.~(\ref{Pi-1}) leads to
\begin{equation}\label{Pi-2}
\hat{\Pi}_{\bm{\pi}}=\int d^3\bm{p} d^3\bm{k}\ \Pi(\bm{k}-\bm{\pi},\bm{p}-\bm{\pi}) |\bm{k}\rangle\langle \bm{p}|,
\end{equation}
where $\Pi(\bm{k},\bm{p})=\langle \bm{k}|\hat{\Pi}_{\bm{0}}|\bm{p}\rangle$.
Then,  Eq.(\ref{X-Trans-inMomentum}) together with Eq.(\ref{Pi-2}), yields
\begin{widetext}
\begin{eqnarray}
\langle \psi_1|\left[\hat{\Pi}_{\bm{\pi}},\hat{\bm{P}}\right]|\psi_2\rangle = \int d^3\bm{p} d^3\bm{k}\ \Pi(\bm{k}-\bm{\pi},\bm{p}-\bm{\pi}) (\bm{p}-\bm{k}) \tilde{\psi}_1^*(\bm{k}) \tilde{\psi}_2(\bm{p})=0, \nonumber \label{Pi-3}
\end{eqnarray}
\end{widetext}
where $|\psi_1\rangle$ and $|\psi_2\rangle$ are arbitrary states in the Hilbert space.
Since linear combinations of functions $\tilde{\psi}_1^*\otimes\tilde{\psi}_2$ are dense in $\mathcal{D}(\mathbb{R}^3\times \mathbb{R}^3)$, Eq.(\ref{Pi-3}) implies 
\begin{equation}
\Pi(\bm{k}-\bm{\pi},\bm{p}-\bm{\pi})=\mathrm{g}(\bm{p}-\bm{\pi})\delta^3(\bm{p}-\bm{k}),
\end{equation}
where $\mathrm{g}(\bm{p}-\bm{\pi})$ is a positive kernel. Therefore 
\begin{equation}\label{Pi-4}
\hat{\Pi}_{\bm{\pi}}=\int d^3\bm{p} \ \mathrm{g}(\bm{p}-\bm{\pi}) |\bm{p}\rangle\langle \bm{p}|.
\end{equation}
Next, the rotational symmetry condition, as given by Eq.~(\ref{Rot-inMomentum}), leads to:
\begin{widetext}
\begin{equation}\label{Pi-5}
\int   \langle \psi|\mathrm{g}(\bm{p}-\bm{\pi})\!\left[|\bm{p}\rangle\langle \bm{p}|,\hat{\bm{J}}\right]\!-i\hbar\bm{\pi}\!\times\! \nabla_{\bm{\pi}} \mathrm{g}(\bm{p}-\bm{\pi})|\psi\rangle d^3\bm{p}=0,
\end{equation}
\end{widetext}
where $|\psi\rangle$ is an arbitrary state. Using the Galilean algebra, by direct calculation, Eq.~(\ref{Pi-5}) simplifies to:
\begin{equation}\label{Pi-6}
\int  (\bm{p}-\bm{\pi})\times\nabla_{\bm{\pi}} \mathrm{g}(\bm{p}-\bm{\pi}) |\tilde{\psi}(\bm{p})|^2  d^3\bm{p}=\bm{0}.
\end{equation}
Since $\tilde{\psi}(\bm{p})$ is an arbitrary wave function, Eq.~(\ref{Pi-6}) implies:
\begin{equation}
(\bm{p}-\bm{\pi})\times\nabla_{\bm{\pi}} \mathrm{g}(\bm{p}-\bm{\pi})=\bm{0},
\end{equation}
This condition ensures that $\mathrm{g}$ is a function of  $|\bm{p}-\bm{\pi}|$ as:
\begin{equation}
 \mathrm{g}(\bm{p}-\bm{\pi})= g(|\bm{p}-\bm{\pi}|).
\end{equation}
Therefore, the momentum POVM is ultimately expressed as:
\begin{equation}\label{Pi-7}
\hat{\Pi}_{\bm{\pi}}=\int d^3\bm{p}\ g(|\bm{p}-\bm{\pi}|) |\bm{p}\rangle\langle \bm{p}|,
\end{equation}
where $g(|\bm{p}-\bm{\pi}|)$ is a radial positive kernel.
\subsection{Uncertainty Relations}
In this subsection, we investigate the uncertainty relations in this extended framework. It is easy to see that, using Eq.(\ref{Pi-7}), Eq.(\ref{Pi-0}) leads to the following expression for momentum uncertainty
\begin{eqnarray} 
(\Delta \pi_i)^2 &=&  \left(\int (\pi_i-\langle \pi_i \rangle)^2 |\tilde{\psi}(\bm{\pi})|^2 d^3\bm{\pi} \right) +\mathsf{p}_0^2, \ \ \ \ \label{Pi-9}
\end{eqnarray}
where $\mathsf{p}_0$ represents the minimal momentum defined as $\mathsf{p}_0= \int g(|\bm{\pi}|) \pi_i^2  d^3\bm{\pi}$.
As shown in the Section \ref{Modified uncertainty relation}, the position uncertainty is given by the following similar expression:
\begin{eqnarray}
(\Delta x_i)^2 &=& \left(\int (x_i-\langle x_i \rangle)^2 |\psi(\bm{x})|^2 d^3\bm{x} \right)+l_0^2. \ \ \ \ \label{Pi-10}
\end{eqnarray}
Moreover, using properties of Fourier transformation, we know that
\begin{widetext}
\begin{eqnarray}\label{Pi-11}
\int (x_i-\langle x_i \rangle)^2 |\psi(\bm{x})|^2d^3x  \int (p_i-\langle p_i \rangle)^2 |\tilde{\psi}(\bm{p})|^2d^3p \geq \hbar^2/4.
\end{eqnarray}
\end{widetext}
Finally, Eqs. (\ref{Pi-9}-\ref{Pi-11}), lead to the following modified position-momentum uncertainty relation:
\begin{eqnarray*}
(\Delta x_i)^2(\Delta \pi_i)^2\geq \frac{\hbar^2}{4}\!+\! l_0^2(\Delta \pi_i)^2\!+\!\mathsf{p}_0^2(\Delta x_i)^2\!+\!l_0^2\mathsf{p}_0^2.\ \
\end{eqnarray*}
It is straightforward to show that by taking the square root, performing a first-order Taylor expansion, and neglecting the final term, $l_0^2\mathsf{p}_0^2$, this relation reduces to the well-known \textit{extended  generalised uncertainty principle},
\begin{eqnarray}\label{Pi-12}
\Delta x_i\Delta \pi_i\geq \! \frac{\hbar}{2}\!\left(\!1\!+\! \frac{2}{\hbar^2}l_0^2(\Delta \pi_i)^2\!+\!\frac{2}{\hbar^2}\mathsf{p}_0^2(\Delta x_i)^2\!\right)\!.\ \ \ \ \
\end{eqnarray}
which has been derived through various approaches in the literature \cite{bolen2005, bambi2008,park2008the,lake2023generalised}. 
A derivation of this relation, similar to our own but with a slightly different interpretation, has been presented in \cite{lake2019generalised}.
\subsection{Dynamics}
In accordance with the general framework of quantum theory, the time evolution of the state vector should be governed by the following Schr\"odinger-type equation:
\begin{equation}
i \hbar \frac{d}{dt} |\psi\rangle = \hat{H} |\psi\rangle,
\end{equation}
For a particle in an external potential $V(\bm{x})$, the Hamiltonian can naturally be defined by   
$\hat H=\hat T+\hat V$,  i.e.,
\begin{equation}
\hat{H}=\int d^3\bm{\pi}\frac{\bm{\pi}^2}{2m}\hat{\Pi}_{\bm{\pi}} +\int d^3\bm{x}V(\bm{x})\hat{Q}_{\bm{x}}.
\end{equation}
Indeed, this Hamiltonian operator constitutes a natural extension of the standard Hamiltonian operator using the following substitution:
\begin{eqnarray*}
|\bm{x}\rangle\langle\bm{x}| &\to & \hat{Q}_{\bm{x}},\\
|\bm{\pi}\rangle\langle\bm{\pi}| &\to & \hat{\Pi}_{\bm{\pi}}.
\end{eqnarray*}
For a free particle, the aforementioned Hamiltonian operator is reduced to 
\begin{equation}
\hat H= \int d^3\bm{\pi}\frac{\bm{\pi}^2}{2m}\hat{\Pi}_{\bm{\pi}}=\frac{\hat{\bm{P}}^2}{2m} +\frac{3\mathsf{p}^2_0}{2m}.
\end{equation}
It is straightforward to verify that the Hamiltonian introduced above satisfies the Galilean algebra. Indeed, the algebra's structure uniquely fixes the form of the free particle Hamiltonian as:
$
\hat H = \frac{\hat{\bm{P}}^2}{2m} +C,
$
in which $C$ is a constant. Interestingly, as we demonstrate below, the structure of non-relativistic dynamics conceptually enforces the condition \( \mathsf{p}_0 = 0 \).
\subsection{Momentum Distribution from Asymptotic Position Distribution}
Based on the quantum theory of measurement, each physical measurement can be described as a position measurement: In principle, the variables that account for the outcome of an experiment are ultimately particle positions \cite{bell2004speakable, holland1995quantum, durr2004quantum, wheeler2014quantum}. This fact has been made clear by John Bell \cite{bell2004speakable}:
\\
\\
 \textit{"In physics the only observations we must consider are position observations, if only the positions of instrument pointers.  ... If you make axioms, rather than definitions and theorems, about the "measurement" of anything else, then you commit redundancy and risk inconsistency."}
\\
\\
 In this regard, in standard non-relativistic quantum mechanics, it is shown that Born's rule for any observable can be derived by considering Born's rule for particle positions \cite{bell2004speakable, holland1995quantum, durr2004quantum}. Specifically, it is shown that the conventional momentum distribution, $|\tilde{\psi}(\bm{p})|^2$, can be derived from the time evolution of the usual position distribution $|\psi(\bm{x})|^2$ \cite{feynman1965quantum, park1968simultaneous,logiurato2006measure}. 
Here, we aim to conduct a similar investigation in our model. To this end, we employ the Feynman-Bohm method to compute the physical momentum distribution through the Time-of-Flight measurements \cite{holland1995quantum, feynman1965quantum, park1968simultaneous,logiurato2006measure,freericks2023measure,ballentine1970statistical}.  
In this method, for measuring the particle's momentum, $\bm{\pi}$,  first we let the initially confined particle fly freely for a considerable amount of time, $t$, then measure its position $\textit{\textbf{x}}$, and set $\bm{\pi}=m\textit{\textbf{v}}$ with $\textit{\textbf{v}}=\textit{\textbf{x}}/t$ 
\footnote{A typical application  of this method would be the determiation of electron momentum in a hydrogen atom by rapid ionization \cite{mccarthy1983real}. For modern applications in atom optics, e.g, see \cite{Bergschneider2019Experimental, Brown2023Time, Wolf2000Ion}.}. 
Therefore the probability of the particle's momentum lying inside the element $d^3\bm{\pi}$ around the point $\bm{\pi}$ is equal to probability of finding the particle's position in the element  $d^3\textit{\textbf{x}}$ around the point $\textit{\textbf{x}}=\textit{\textbf{v}}t$
\footnote{An inaccuracy stems from the unknown initial position $\textit{\textbf{x}}_0$, as the correct formula would be $\textit{\textbf{x}}=\textit{\textbf{v}}t+\textit{\textbf{x}}_0$, but in the limit $t \to \infty$ the contribution from $\textit{\textbf{x}}_0$ vanishes \cite{logiurato2006measure}.}; i.e. 
$
\tilde{\rho}(\bm{\pi})d^3\bm{\pi}=\lim_{t\to\infty}[\rho(\textit{\textbf{x}},t)d^3\textit{\textbf{x}}]_{\textit{\textbf{x}}=\textit{\textbf{v}}t},
$
where $\tilde{\rho}(\bm{\pi})$ represents the momentum probability density. Explicitly, using $\bm{v}=\bm{\pi}/m$, we have  
\begin{equation} \label{Momentum-From-TOF}
\tilde{\rho}(\bm{\pi})= \frac{1}{m^3} \lim_{t\to\infty}t^3\rho(\bm{\pi}t/m,t),
\end{equation}
where $1/m^3$ is Jacobian of the $(\textit{\textbf{v}} \to \bm{\pi})$ transformation. Using the time-dependent position probability density in our model,
\begin{widetext}
$$
\rho (\bm{x},t)\!=\!\!\int \! d^3\bm{k}'d^3\bm{k}'' f(|\bm{k}'-\bm{k}''|) \tilde{\psi}_t(\bm{k}')\tilde{\psi}_t^*(\bm{k}'') e^{\frac{i}{\hbar}(\bm{k}'-\bm{k}'').\bm{x}},
$$
\end{widetext}
Eq.(\ref{Momentum-From-TOF}) leads to
\begin{equation}
\rho(\bm{\pi}t/m,t){=}\int A(\textit{\textbf{k}}',\textit{\textbf{k}}'') e^{itB_{\bm{\pi}}(\textit{\textbf{k}}',\textit{\textbf{k}}'')} d^3\textit{\textbf{k}}' d^3\textit{\textbf{k}}'',
\end{equation}
where
\begin{equation}
A(\textit{\textbf{k}}',\textit{\textbf{k}}'')\equiv\tilde\psi(\textit{\textbf{k}}')\tilde\psi^*(\textit{\textbf{k}}'')f(|\textit{\textbf{k}}'-\textit{\textbf{k}}''|),
\end{equation}
and
\begin{equation}
B_{\bm{\pi}}(\textit{\textbf{k}}',\textit{\textbf{k}}'')\equiv\frac{(\textit{\textbf{k}}'-\textit{\textbf{k}}'').{\bm{\pi}}}{m\hbar}\!-\!(\frac{\bm{k}'^2-\bm{k}''^2}{2m\hbar}).
\end{equation}
Next, using the stationary phase method \cite{mcclure1990multidimensional,fedoryuk1971stationary}, we get
\begin{widetext}
$$\lim_{t\to\infty}\!t^3\rho({\bm{\pi}}t/m,t)=(2\pi)^3A({\bm{\pi}},{\bm{\pi}})|\det \mathcal{H}_B|^{-\frac{1}{2}} e^{i\pi \sign(\mathcal{H}_B)/4},$$
\end{widetext}
where $\mathcal{H}_B$ is Hessian matrix of $B_{\bm{\pi}}(\textit{\textbf{k}}',\textit{\textbf{k}}'')$ at the critical point (i.e. $\textit{\textbf{k}}'=\textit{\textbf{k}}''=\bm{\pi}$, where $\nabla_{\textit{\textbf{k}}'}B=\nabla_{\textit{\textbf{k}}''}B=0$), and $\sign(\mathcal{H}_B)$ is signature of it.  By direct calculation, it is easy to see that $|\det \mathcal{H}_B|=(m\hbar)^{-6}$ and $\sign(\mathcal{H}_B)=0$, which beside the fact that $f(0)= (2\pi\hbar)^{-3}$,  lead to standard momentum distribution,
\begin{equation}
\tilde{\rho}(\bm{\pi})=|\tilde\psi(\bm{\pi})|^2. 
\end{equation}
This momentum distribution imply that 
\begin{equation}
(\Delta \pi_i)^2 =  \left(\int (\pi_i-\langle \pi_i \rangle)^2 |\tilde{\psi}(\bm{\pi})|^2 d^3\bm{\pi} \right), 
\end{equation}
which, when compared with Eq. (\ref{Pi-9}), leads to $\mathsf{p}_0 = 0$. This implies that the proposed generalization reduces to the model presented in the manuscript.

In summary, ensuring consistency with Galilean symmetry requires that the minimum momentum uncertainty be zero. This result is not entirely surprising, as derivations of extended uncertainty relations (with non-zero minimum momentum uncertainty) are typically conducted in spacetimes that correspond to the non–zero cosmological constant, such as (Anti-)de Sitter spacetime \cite{bolen2005, bambi2008, park2008the, mignemi2010,lake2019generalised}, rather than in non-relativistic spacetimes with Galilean symmetry. Breaking Galilean symmetry may permit a non-zero minimum momentum uncertainty; however, such an analysis is beyond the scope of this paper.

\bibliographystyle{quantum}
\bibliography{mybibliography}

\end{document}